\documentclass[12pt,twocolumn]{iopart}
\usepackage{iopams}  
\usepackage{graphicx}

\def \musr {$\mu^+$SR}
\def \fmu {F--$\mu$}
\def \fmuf {F--$\mu$--F}
\def \lacuo {La$_2$CuO$_4$}
\newcommand{\text}[1]{\textrm{#1}}

\begin{document}

\comment[Localizing muon sites]{Playing quantum hide-and-seek with the muon: localizing muon stopping sites}

\author{J.~S.~M\"oller,$^{1}$ P.~Bonf\`a,$^2$ D.~Ceresoli,$^3$ F.~Bernardini,$^4$ S.~J.~Blundell,$^1$ T.~Lancaster,$^5$ R.~De~Renzi,$^2$ N.~Marzari,$^6$ I.~Watanabe,$^7$ S. Sulaiman,$^8$ M. I. Mohamed-Ibrahim.$^8$}
\address{$^1$Department of Physics, Oxford University, Parks Road, Oxford, OX1 3PU, UK; $^2$Dipartimento di Fisica e Scienze della Terra and Unit\`a CNISM di Parma, Universit\`a di Parma, I-43124 Parma, Italy; $^3$Istituto di Scienze e Tecnologie Molecolari CNR, via Golgi 19, 20133 Milano, Italy; $^4$CNR-IOM-Cagliari and Dipartimento di Fisica, Universit\`a di Cagliari, IT-09042 Monserrato, Italy; $^5$Centre for Materials Physics, Durham University, South Road, Durham DH1 3LE, UK; $^6$Theory and Simulation of Materials (THEOS), \'{E}cole Polytechnique F\'ed\'erale de Lausanne, 1015 Lausanne, Switzerland, $^7$Advanced Meson Science Laboratory, RIKEN Nishina Center, 2-1 Hirosawa, Wako, Saitama 351-0198, Japan; $^8$Computational Chemistry and Physics Laboratory, School of Distance Education, Universiti Sains Malaysia, 11800 Penang, Malaysia.}
\ead{johannes.moeller@physics.ox.ac.uk}

\begin{abstract}
One of the most fundamental limitations of a muon-spin relaxation experiment can be the lack of knowledge of the implantation site of the muon and the uncertainty about the muon's perturbation of its host. Here we review some of the work done on the `muon site problem' in the solid state and highlight some recent applications of electronic structure calculations that have successfully characterized the quantum states of muons in a number of insulating compounds containing fluorine, in a number of pnictide superconductors, and in ZnO.
\end{abstract}

\maketitle

\section{Introduction}
A muon-spin relaxation (\musr) experiment involves implanting spin-polarized positive muons in a sample in order to probe the local static and dynamic magnetic properties (a more detailed introduction can be found in Ref.~\cite{Blundell99} and in the lead article of this series). \musr\ is an extremely sensitive probe of magnetism (for an illustration see, for example, Ref.~\cite{Pratt2011Nature}) but it has two significant limitations. The first concerns the lack of knowledge of the site of implantation of the muon, which hinders the measurement of magnetic moments or the comparison of different candidate magnetic structures using \musr. Second, the unknown extent of the perturbation due to the muon of the local crystal and electronic structure of the host has been the cause for increased concern since \musr\ is frequently employed in the study of systems that lie on the verge of ordering~\cite{Pratt2011Nature,Kojima1995PRL,Blundell1997JPhysCM,Lancaster2006PRB} or where doping is a critical parameter~\cite{Luke1990PRB,Amit2010PRB,Jack2012PRB}. 

From the very beginning of \musr\ significant effort has been devoted to the determination of muon sites. In some materials a determination of interstitial muon sites was indeed possible thanks to accurate experimental studies of the Knight shift~\cite{Renzi1984Location}, level crossing resonances~\cite{Kiefl1988,Brewer1991}, by inspecting relaxation rates as a function of applied field~\cite{Camani77,Chow1994}, or through the observation of quantum entanglement between the muon spin and a small number of surrounding nuclei (discussed in more detail below)~\cite{brewer86PRB,Lancaster2007PRL}. Nonetheless the number of examples where the muon site can be determined by experimental means alone is limited and even in those cases the experimental information about the muon site and the perturbation caused by the muon is usually incomplete. An improved understanding of the muon state in solids would not only benefit a more complete understanding of the nature of the muon response in a wide number of compounds, it could also enable a determination of magnetic moments and perhaps even allow to differentiate between different models of magnetic structures. This information would be particularly valuable in a number of topical compounds where the observation of magnetic neutron scattering is challenging, such as compounds containing nuclei that strongly absorb neutrons (for example iridates) or compounds with particularly small magnetic moments (for example frustrated and low-dimensional systems, where the moments are strongly renormalized by fluctuations).

In this Comment we present three case studies that characterize the muon states in solids using {\it ab initio} electronic structure theory. Previous work in this area has focussed on the paramagnetic states formed by muons and protons in semiconductors, for a review see~\cite{Cox2009}. Diamagnetic muon states (where the contact hyperfine coupling is negligible) have received very little attention in spite of their greater utility in the study of magnetic materials. {\it Ab initio} methods have also been applied in the study muoniated molecular radicals, which is the subject of another Comment in this series. Here we summarize a number of recent applications of density-functional theory~(DFT) that focus on the diamagnetic muon states in a number of solids. 

\section{Quantum states of muons in insulating fluorides}
In host compounds containing fluorine, diamagnetic muons can couple strongly to the fluoride ions, often forming linear \fmuf\ complexes~\cite{brewer86PRB}, although bent \fmuf\ and \fmu\ geometries have been shown to exist as well~\cite{Lancaster2007PRL}. The magnetic dipolar coupling between muon and fluorine nuclear spins $I$ (both $I=1/2$) is described by the Hamiltonian 

\begin{equation}
\label{eq:hamiltonian}
\mathcal{H} = \sum_{i > j} \frac{\mu_{0} \gamma_{i} \gamma_{j}}{4 \pi
  r^{3}} 
\left[ \mathbf{I}_{i} \cdot \mathbf{I}_{j} - 
3 (\mathbf{I}_{i} \cdot \hat{\mathbf{r}})(\mathbf{I}_{j} \cdot \hat{\mathbf{r}}  ) \right], 
\end{equation}
where $\hat{\mathbf{r}}$ is the normalized vector connecting spins $i$ and $j$, $\gamma_i$ is the gyromagnetic ratio of spin $i$, $r$ is the distance between spins $i$ and $j$ and all other symbols take their usual meaning. This interaction gives rise to a characteristic muon precession signal (which can be easily determined by diagonalizing the Hamiltonian) that is sensitive to the geometry of the muon-fluorine state, allowing an accurate experimental determination of the muon's local site geometry~\cite{brewer86PRB,Lancaster2007PRL}. Two recent studies~\cite{Moeller2013PRBR,Bernardini2013} have investigated the quantum states of muons in the non-magnetic ionic insulators LiF and NaF (rock-salt structure), CaF$_2$ and BaF$_2$ (fluorite structure), YF$_3$ (orthorhombic), and for the antiferromagnetic insulator CoF$_2$ (rutile-type structure). 

\begin{figure}[htbp]
\includegraphics[width=0.45\textwidth]{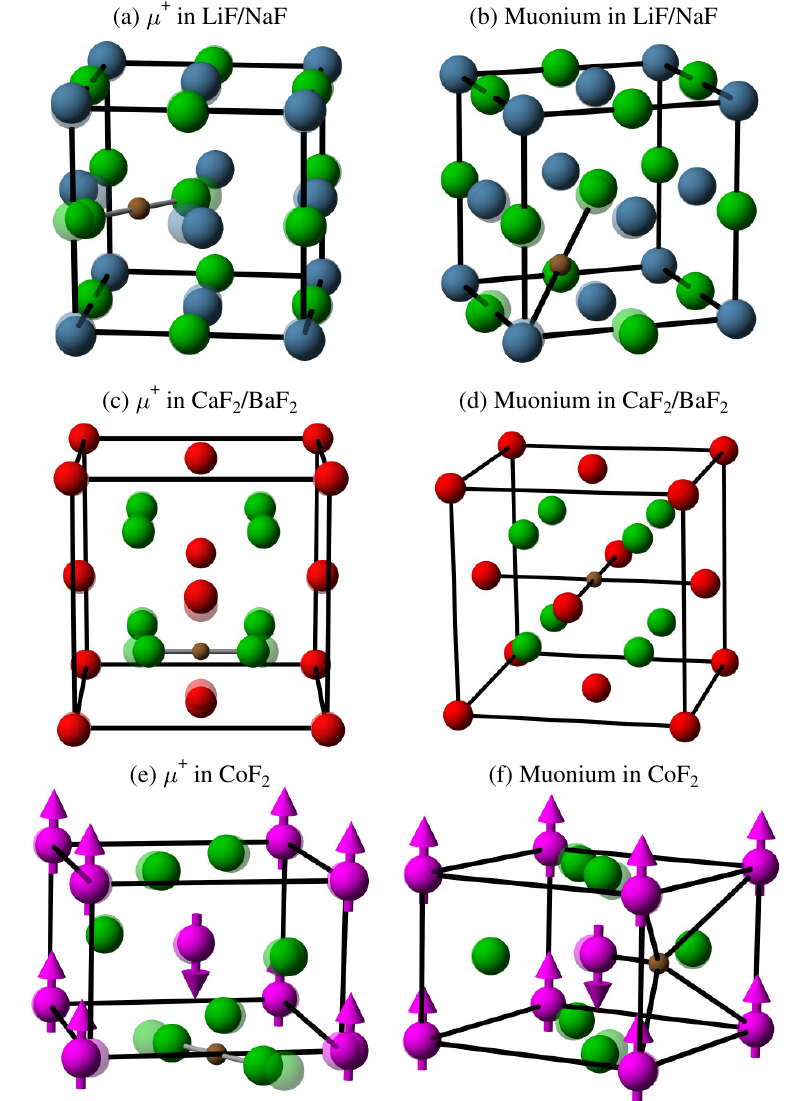}
\caption{\label{fig:structures}(Color online). Calculated equilibrium geometries of dia- and paramagnetic muon states in LiF/NaF (Li/Na blue, F green), CaF$_2$/BaF$_2$ (Ca/Ba red), and CoF$_2$ (Co magenta). Translucent spheres represent the equilibrium ionic positions before the muon (brown) is introduced into the crystal. Black lines are a guide to the eye. The $c$~axis is vertical. From Ref.~\cite{Moeller2013PRBR}.}
\end{figure}

\begin{figure} \includegraphics[width=0.45\textwidth]{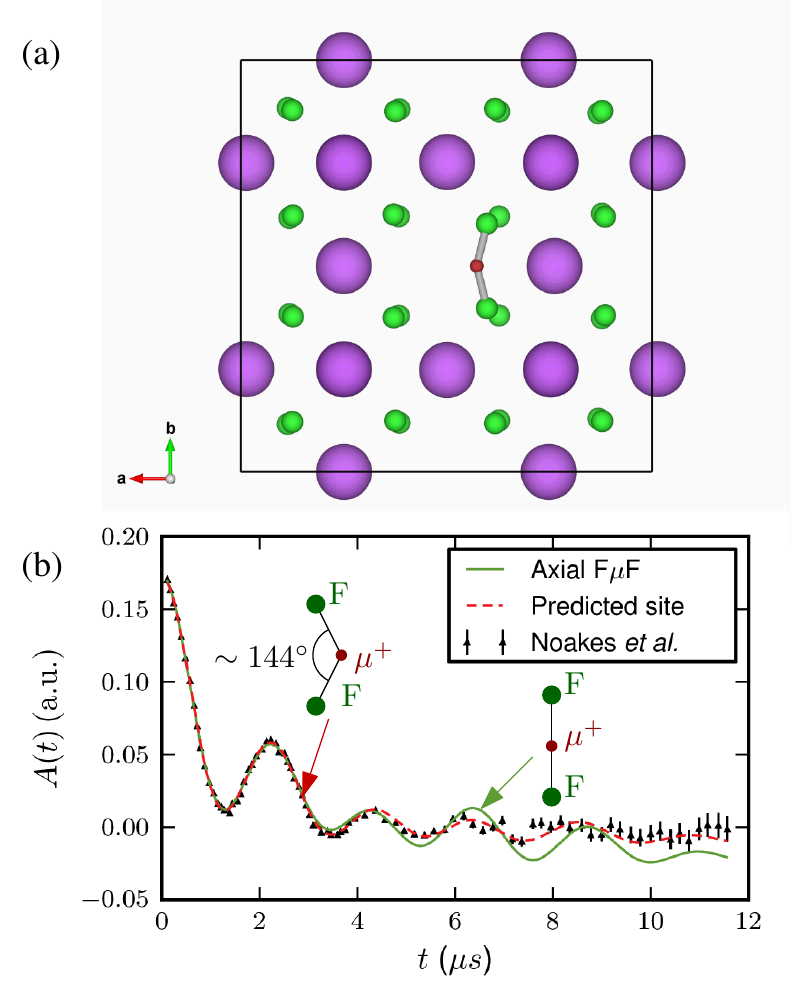} 
\caption{(Color online) (a) The crystal 
structure of YF$_{3}$ with muon after structural relaxation. Y(purple), F (green), muon (dark red). The positions of the Y atoms are only marginally affected by the interstitial muon. (b) Comparison between the analysis of Noakes {\it et al.}~with a linear F-$\mu$-F configuration (Ref. \cite{Noakes1993}) and the results obtained with the procedure outlined in the text which predicts a slightly distorted bond between the muon and the nearest neighbour F atoms. Based on figures in Ref.~\cite{Bernardini2013}. The data were visualized with {\it VESTA}~\cite{Momma2011}.\label{fig:yf3}}
\end{figure}

Both studies used the plane-wave pseudopotential method as implemented in the {\sc Quantum ESPRESSO} package~\cite{qespresso}. These calculations employ periodic boundary conditions and so to reduce the error due to defects in neighbouring unit cells, a supercell approach was used where each supercell contained $2\times2\times2$ conventional unit cells (except for YF$_3$ where the conventional orthorhombic unit cell was used). In these studies, the charge state of the muon was determined by the charge of the supercell (+1 for diamagnetic and neutral for paramagnetic states). Two alternative approaches were used for determining the relaxed geometries: the first placed a muon in several randomly chosen low-symmetry sites and all ions were allowed to relax until the forces on all ions and the energy change between iterations had fallen below a convergence threshold~\cite{Moeller2013PRBR}; the second calculated the electrostatic potential of the unperturbed solid first, placed muons in the local minima of the electrostatic potential, and then the structure was allowed to relax~\cite{Bernardini2013}. 

Figs.~\ref{fig:structures} and \ref{fig:yf3} show the calculated equilibrium geometries of the muon states in the compounds considered. In all cases, an \fmuf\ state is the lowest energy state. In LiF, NaF, CaF$_2$, and BaF$_2$ the calculations correctly predict the experimentally known geometries~\cite{brewer86PRB} with great accuracy: bond lengths are within $\sim 3\%$ of the experimental values. Even though the muon site in CoF$_2$ agrees with the site known from a detailed experimental study~\cite{Renzi1984Location}, the predicted \fmuf\ state had not been observed experimentally. Following their {\it ab initio} work, the authors experimentally searched for signatures of an \fmuf\ state in CoF$_2$ and found unambiguous evidence for a linear \fmuf\ state of a geometry that is in excellent agreement with their DFT prediction~\cite{Moeller2013PRBR}. In YF$_3$ an \fmuf\ signal had been previously observed~\cite{Noakes1993} and was attributed to the formation of a linear \fmuf\ state. In their DFT calculations, the authors found several candidate sites for the muon. On kinetic grounds they predict that the ground state diamagnetic site of the muon in YF$_3$ is instead a bent \fmuf\ state with a bond angle of about 144$^\circ$. They revisit the previous experimental data and show that the geometry obtained from {\it ab initio} calculations is indeed in better agreement with the experimental data than the previously suggested linear \fmuf\ state (see Fig.~\ref{fig:yf3})~\cite{Bernardini2013}. All of these results demonstrate the accuracy with which muon sites can be determined in insulators. 

Based on their calculated structures, the authors of Ref.~\cite{Moeller2013PRBR} also study the distortions introduced by the muon. They show that the crystallographic distortions are  significant at short range, with nearest neighbour (n.n.) distortions of up to 0.5~\AA. While it was known that the perturbation of the fluoride ions must be significant based on the experimentally measured \fmu\ bond lengths of the \fmuf\ states found in many fluorides, these calculations allow the cation distortions to be quantified as well. Since localized magnetic moments would be located on the cation, the cation displacements are particularly pertinent to understanding the effect of the muon's perturbation on experimentally measured \musr\ spectra. The authors demonstrate that in the perturbation of the n.n.\ cations can even exceed those of the fluoride ions bound in the \fmuf\ state. In antiferromagnetic CoF$_2$ they calculate that this will lead to a reduction of the observed muon precession frequency by just over 20\%, in good agreement with an estimate based on experimental data~\cite{Renzi1984Location} of 16\%. This correction should be taken into account if magnetic moments were to be measured accurately in an ionic insulator. Since at short distances the distortions are mainly caused by the electrostatic interaction of the unscreened muon with its surroundings, the authors expect similar distortions in any ionic insulator, while the muon is likely to be more screened in more covalent compounds, probably leading to smaller distortions. 

\begin{figure*}[htbp]
\includegraphics[width=0.95\textwidth]{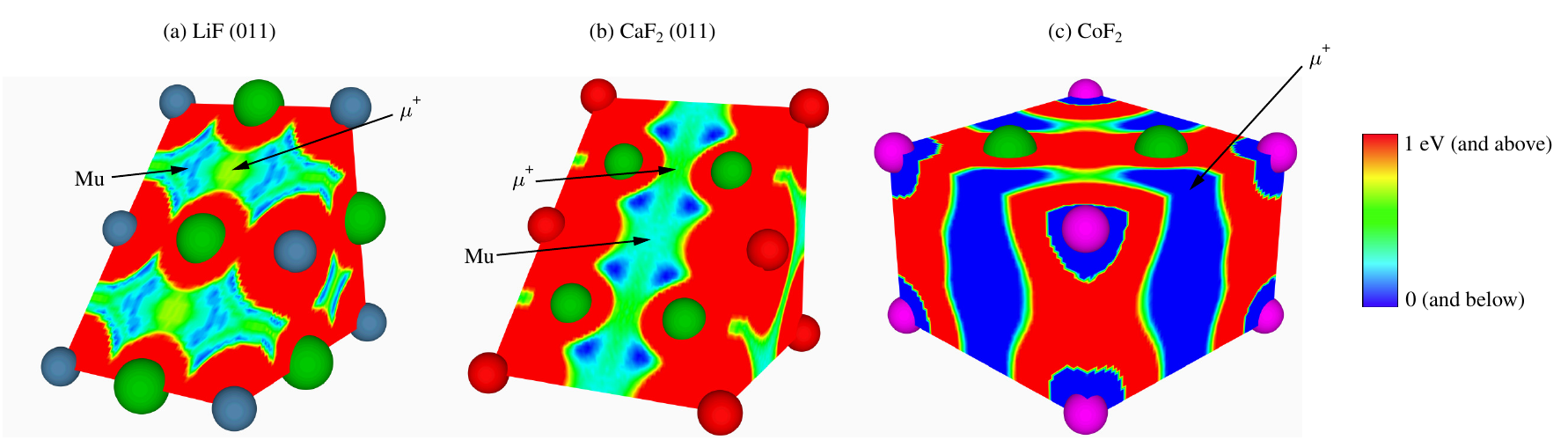}
\caption{\label{fig:potential}(Color online). Calculated electrostatic potential for the unperturbed solid. Blue coloring indicates regions that are attractive to a positive charge, red regions repel a positive charge. Below and above the end of the scale the color coding is blue and red, respectively, with no further gradient. The scale is relative and cannot be compared between different compounds. Ions are drawn at their ionic radii. Li (blue), F (green), Ca (red), Co (magenta). The c axis is vertical. Arrows indicate the dia- and paramagnetic muon sites obtained through a full relaxation, which agree with the experimentally determined muon sites. In CoF$_2$ the muonium site is close to the octahedral site that also hosts the diamagnetic muon. The muon zero point energy, characterizing the extent of its delocalization in the absence of bonding, is about 0.8~eV in the \fmuf\ state and about $0.2-0.6$~eV as muonium. The data were visualized with {\it VESTA}~\cite{Momma2011}. From Ref.~\cite{Moeller2013PRBR}.}
\end{figure*}

The quantum nature of nuclei is generally ignored in DFT calculations since nuclear masses are typically so large that quantum effects (caused for example by the spread of the nuclear wavefunction) lie below the current level of accuracy of the technique. However, at approximately $1/9$ the mass of a proton the muon is an exceptionally light impurity. Quantum effects can therefore be expected to play a more significant role in the localization of a muon than for the majority of conceivable point defects. The quantum properties of the muon in the \fmuf\ state were estimated using density-functional perturbation theory to calculate the vibrational properties of the \fmuf\ molecule. The zero-point energy (ZPE) was then estimated from the calculated vibrational frequencies in the harmonic approximation. This approximation neglects the finite spread of the muon wavefunction and anharmonic terms in the potential. However, it takes account of the coupled muon-ion zero-point motion and is most appropriate for a \emph{molecular defect} such as the \fmuf\ system.  In fact it was found that in most of these systems the vibrational modes of the \fmuf\ centre decouple from the rest of the crystal and so the \fmuf\ system can be viewed as a molecule-in-a-crystal defect. The F--$\mu$ bond is the strongest known hydrogen bond in nature and combined with the small muon mass this leads to the \fmuf\ centre possessing an exceptionally large ZPE: larger than that of any natural triatomic molecule~\cite{Moeller2013PRBR}. This demonstrates the importance of quantum effects on muon localization.

There has been considerable interest recently in identifying muon sites by locating the minima of the electrostatic potential of the unperturbed host~\cite{Luetkens2008,Maeter2009PRB,Bendele2012,deRenzi2012,Prando2013}. The authors of Refs.~\cite{Moeller2013PRBR,Bernardini2013} have therefore compared the muons sites in this series with the location of the minima of the electrostatic potential of the unperturbed solid, and have found that these do not generally coincide (see Fig.~\ref{fig:potential}). In the diamagnetic case this is primarily due to the formation of the molecular \fmuf\ state. All of the compounds studied here are very ionic in character and the $\mu^+$-lattice interaction is therefore expected to be stronger than in more covalent insulators or metals, where the $\mu^+$ charge would at least be partially screened. However, if the muon charge were completely screened there would be no reason why a muon should localize in an electrostatic minimum. We expect the combination of this screening (where operative), the muon-lattice interaction, and the muon's exceptionally large zero-point energy to frequently lead to muon localization away from the minima of the electrostatic potential of the unperturbed host. We therefore believe that muon sites cannot be determined reliably on the basis of the electrostatic potential alone. 

\section{Muon sites in metallic systems: pnictide superconductors}
In this section we discuss the application of DFT to the determination of muon sites in pnictide superconductors. The analysis of the magnetic ground state properties by means of \musr\ in the pnictides triggered the interest to calculate the muon stopping sites~\cite{Maeter2009PRB,deRenzi2012,Lamura2013}. It is well known that the magnetic and structural properties of pnictides are not accurately described in DFT because of electronic correlations. Nonetheless we have demonstrated that it is possible to determine muon sites that are consistent with the experimental \musr\ data.

\begin{figure}
\includegraphics[width=0.45\textwidth]{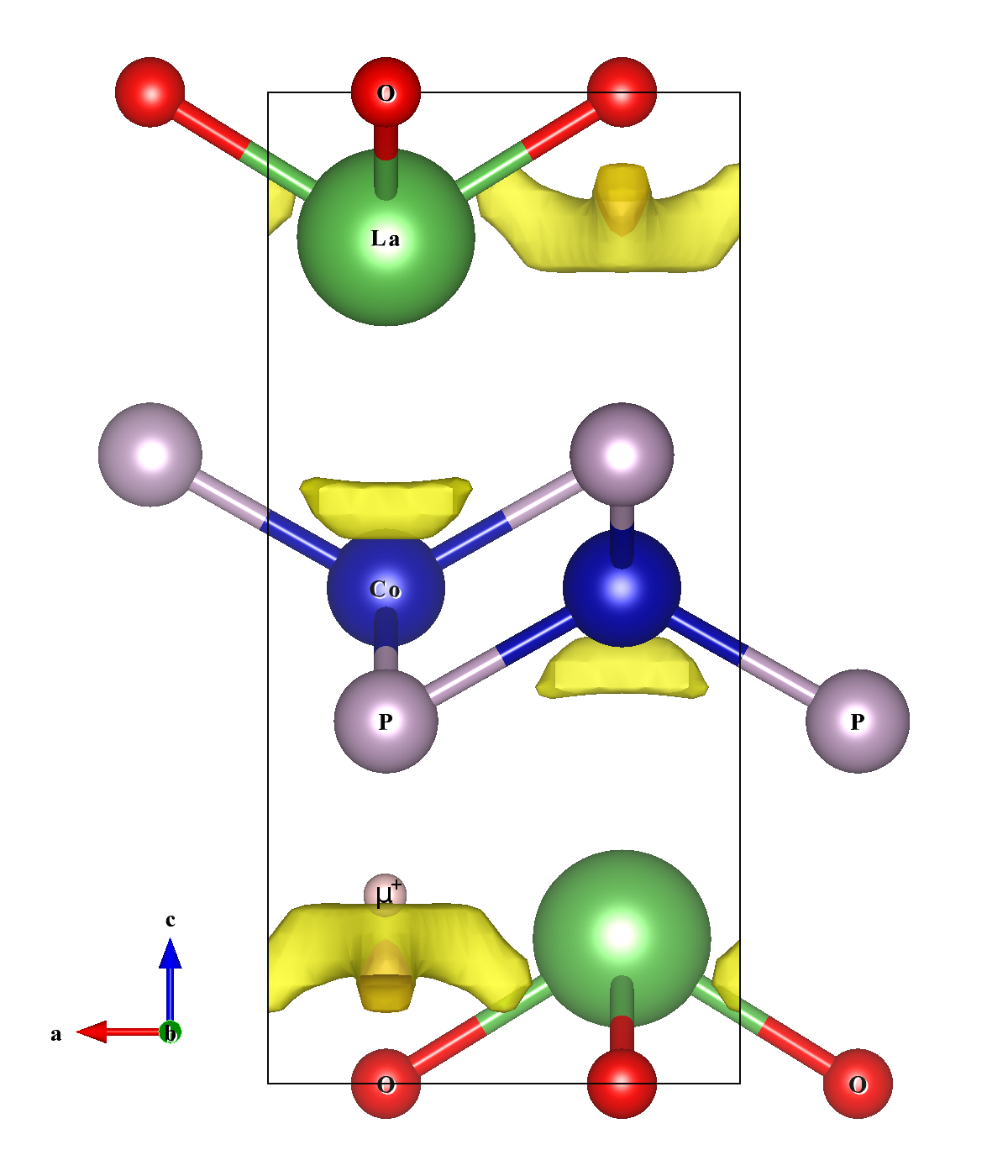}\label{fig:lacopo} 
\caption{(Color online) Comparison of the localization volume based on the electrostatic potential of the unperturbed host (shaded yellow) and the relaxed muon site obtained through a structural relaxation (labelled $\mu^{+}$) in LaCoPO. Further details can be found in Ref.~\cite{Prando2013}.}    
\end{figure}

Muon sites were identified by calculating the electrostatic potential of the unperturbed host. In some cases the muon was then allowed to relax starting from a local minimum in the electrostatic potential using a neutral supercell. While the muon site does not coincide with the local minima in the electrostatic potential for the fluorides studied above~\cite{Moeller2013PRBR,Bernardini2013}, the muon charge is screened in these metallic systems, preventing strong bonding and so in these system this is a better approximation (see Fig.~\ref{fig:lacopo}). The correct evaluation of the muon zero-point motion is a key factor in muon position evaluation. Indeed many interstitial sites that might be stable sites for a heavier particle (for example a proton) are not stable for the muon. Bernardini {\it et al.}~ introduced the concept of the \emph{localization volume}~\cite{Bernardini2013} as the volume defined by the potential isosurface $V(r)=E_{0}$ where $V(r)$ is the electrostatic potential and $E_{0}$ is the ground state energy for the muon in the electrostatic potential. 

The ground state energy $E_{0}$ was found by solving the Schr\"odinger equation for the muon in the electrostatic potential of the host (either for the unperturbed host or with relaxed ionic positions due to the presence of the muon). This `rigid-lattice' approximation takes full account of anharmonic terms in the potential, but it neglects the effect of the muon on the surrounding charge density and the coupled muon-ion zero-point motion. Its greatest advantage is the much reduced complexity of the calculation compared with calculating the vibrational modes, as described above, which is approximately a factor 3$N$ more computationally expensive ($N$ being the number of atoms in the supercell). It is also possible to use the total energy from a series of self-consistent calculations including the muon with different muon positions as potential for which the Schr\"odinger equation is solved. This would take account of the muon's effect on the surrounding charge density, but would be costly to do for a full three-dimensional grid. In either case this approximation is most appropriate for an \emph{atomic defect} such as muonium~\cite{Moeller2013PRBR} or a diamagnetic muon in a screened environment such as a metal~\cite{Lamura2013,Prando2013,deRenzi2012}, as is the case here.

A number of recent successes~\cite{Lamura2013,Prando2013,deRenzi2012} with pnictides demonstrate that the prediction based on the solution of the muon Schr\"odinger equation for the unperturbed lattice can be as accurate as required to understand and extract quantitatively consistent results from \musr\ spectra. A few representative cases are reported in Table~\ref{tab:comp}. In Fig.~\ref{fig:lacopo} we compare the muon position in LaCoPO~\cite{Prando2013} estimated by considering the minimum of the electrostatic potential with the one obtained through a full ionic relaxation (including the muon) the procedure outlined above: the displacement from the potential minimum is approximately 0.25 \AA. This displacement can have a significant effect on the calculated dipolar field at the $\mu^{+}$ site.

\begin{table}
\center
    \begin{tabular}{ccc}
    Compound & $\mathbf{B}_{\rm dip}^{\rm calc}$ [G] & $\mathbf{B}_{\rm dip}^{\rm exp}$ [G] \\
    \hline
    FeTe    & 2230 & 2000(100) \\
    LaFeAsO site 1 & 1530 & 1650(50) \\
    LaFeAsO site 2 & 270 & 180(10) \\
\end{tabular}
\caption{Comparison of the experimental ($\mathbf{B}_{dip}^{exp}$) 
and the calculated ($\mathbf{B}_{dip}^{calc}$) dipolar field at the $\mu^{+}$ site for two parent compounds of the pnictide high-temperature superconductors. The minima of the electrostatic potential are used to evaluate $\mathbf{B}_{dip}^{calc}$ on the basis of the neutron scattering results for the Fe magnetic moments reported in Refs. \cite{Martinelli2010,Qureshi2010} (only significant figures are reported). \musr\ results are taken from Refs. \cite{Lamura2013,deRenzi2012} which contains further details of the calculation.\label{tab:comp}}
\end{table}

\section{MO cluster and potential methods to determine the muon site and hyperfine interactions in \lacuo\ and ZnO}
Even in \lacuo, the parent compound of the family with the simplest crystal structure of high-T$_{\rm C}$ compounds, the exact muon stopping site is uncertain. Based on dipole-field calculations, Hitti {\it et al.}~\cite{Hitti1990} have estimated the muon stopping site to be near the apical oxygen of the CuO$_6$ octahedra in \lacuo, which was supported by later calculations of the electrostatic potential~\cite{Nachumi1998}. A different study~\cite{Torikai1993} suggested the muon position to be 1~\AA\ away from the in-plane oxygen through measurements of the nuclear dipole field distribution at the muon site, aided by a calculation of the electrostatic potential. Such a stable binding state between oxygen and muon has also been suggested by other {\it ab initio} calculations~\cite{Sulaiman1994}. More recent work has lead to even more suggestions about the muon stopping site in \lacuo~\cite{Suter2003,Huang2012}.

\begin{figure}
\includegraphics[width=0.26\textwidth]{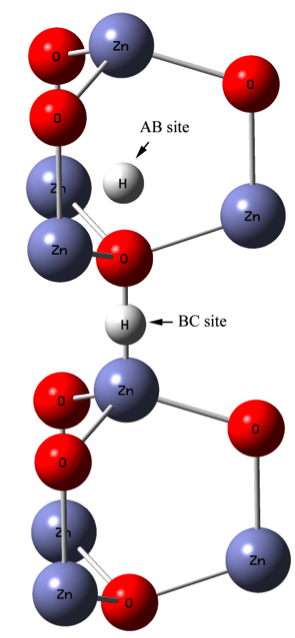}\label{fig:zno} 
\caption{(Color online) The Anti Bonding (AB) and Bond Center (BC) sites for muonium in ZnO~\cite{Watanabe_prep}.}
\end{figure}

In order to shed more light on this problem, Watanabe {\it et al.}~\cite{Watanabe_prep} propose to develop a strategy for identifying muon stopping sites by studying the muon sites in ZnO, where some information about the muon sites is available experimentally. ZnO is a wide-gap semiconductor that has been extensively studied due to its technological significance~\cite{VdWZnO}. ZnO tends to exhibit n-type conductivity, although the source of this conductivity remains controversial. Two \musr\ studies confirmed the existence of muonium centers in ZnO; one experiment observed only a single muonium center~\cite{Cox2001} with contact and dipolar hyperfine couplings of $A=500\pm20$~kHz and $D=260\pm20$~kHz, respectively, while the other study detected two signals corresponding to two distinct muonium centers~\cite{Shimomura2001}. These two centers were proposed to be the so-called Anti-Bonding center (AB) with $A=491(5)$~kHz and $D=265(9)$~kHz and the Bond Center (BC) site with $A = 293(7)$~kHz and $D=286(13)$~kHz (see Fig.~\ref{fig:zno}). 

For ionic compounds, the customary practice to treat the boundary conditions is to embed the cluster with a finite number of point charges that would reproduce the correct Madelung potentials~\cite{Mitchell1991}. ZnO has both ionic and covalent character in its bonding. The usage of hydrogen to terminate the dangling bonds could therefore have significant effects on the electronic structure, especially for the BC site. To examine the effects of hydrogen terminating dangling bonds in the molecular-orbital (MO) cluster method, Watanabe {\it et al.}~have performed MO cluster calculations for muonium at the BC site with and without the hydrogen terminators. The clusters contained eight Zn and O atoms and one hydrogen to represent the muonium. For muonium at the BC site, Watanabe {\it et al.}~found that the lattice relaxation effect is about 40\% which is consistent with a previous {\it ab initio} study~\cite{VdWZnO}. 

Watanabe {\it et al.}~have also employed both Hartree-Fock and density-functional theory calculations (the PBE and B3LYP functionals were used with similar results) to calculate the hyperfine coupling constants for muonium at the BC site using the Gaussian 03 software. They found that by not using hydrogen as terminators, the hyperfine coupling constants were reduced significantly. Comparing to the results of H. Li {\it et al.}~\cite{Li2003}, the isotropic Fermi contact coupling constant $A$ is reduced by a factor of 35 while for the dipolar component $D$ the reduction is by a factor of 56. For $A$, the sign remained negative both for clusters with and without hydrogen terminators. The hyperfine coupling constants calculated using DFT were smaller than those obtained previously~\cite{Li2003} and the sign of $A$ was positive in Watanabe {\it et al.}'s work, in agreement with the experimental data and improving on previous work~\cite{Li2003}. Further work will investigate whether embedding the cluster in an assembly of point charges would further bring the values of the hyperfine coupling constants closer to the experimental ones.

\section{Conclusions}
We have discussed the motivation for investigating the location of muon sites and the extent of the perturbation caused by the muon. We have presented recent successes in the study of muon states in wide-gap insulating fluorides, where the local muon site can be determined experimentally with high accuracy, in pnictide superconductors and in ZnO. These results demonstrate that DFT is a powerful tool to characterize muon states in a wide range of solids. With the continuing improvements of electronic-structure methods and the growing performance of the computational resources available, muon states can be explored more accurately and in greater detail than ever before, even in challenging materials. We believe that this will become a routine part of many muon experiments and will boost the range of physical properties that can be explored with \musr. This work is supported by EPSRC (UK), the European 7$^{\rm th}$ framework programme contract 226507 (NMI3), RIKEN (Japan), and Universiti Sains Malaysia. \\

\bibliographystyle{unsrt}
\bibliography{/Users/jsm/dphil/literature/Library}

\end{document}